\documentclass[12pt]{article}

\usepackage{epsfig,amsfonts}

\newcommand{\betac}{\beta_\mathrm{c}}

\def \be {\begin{equation}}
\def \ee {\end{equation}}                                         
\def \ba {\begin{eqnarray}}
\def \ea {\end{eqnarray}}

\begin{document}

\title{First order signatures in 4D pure compact
U(1) gauge theory with toroidal and spherical topologies.}

\author{I. Campos, A. Cruz and A. Taranc\'on.}
\bigskip
\maketitle

\begin{center}
{\it Departamento de F\'{\i}sica Te\'orica, Facultad de Ciencias,\\
Universidad de Zaragoza, 50009 Zaragoza, Spain \\
\small e-mail: \tt isabel, tarancon, cruz@sol.unizar.es} \\
\end{center}
\bigskip

\begin{abstract}

We study the pure compact U(1) gauge theory with the
extended Wilson action ($\beta$, $\gamma$ couplings) by finite
size scaling techniques, in lattices ranging from $L$=6 to $L$=24
in the region of $\gamma\leq 0$ with toroidal and spherical topologies.
The phase transition
presents a double peak structure which survives in the thermodynamical
limit in the torus. 
In the sphere the evidence supports the idea of a weaker, but still 
first order, phase transition. For $\gamma<0$ the 
transition becomes weaker and larger lattices are needed to find its
asymptotic behaviour. Along the transient region the behaviour is the 
typical one of a weak first order transition for both topologies, 
with a region where $1/d<\nu<0.5$,
which becomes $\nu\approx 1/4$ when larger lattices are used.

\end{abstract}

\newpage

\section{Introduction}

The four dimensional pure compact U(1) gauge theory is the simplest gauge
abelian interaction we can describe in the lattice. This model
is known to posses a phase transition (PT) line separating
a Confined phase from a Coulomb one. Big efforts have been devoted
to the study of the order of this PT, which turned out to be
a controversial issue. The implementation of pure compact U(1) in the lattice
with the Wilson action:
\be
S_{\rm W} = - \sum_{P} \beta \cos \theta_{\rm P}
\label{WIL}
\ee
has been studied for a long time. At first, in the eighties,
the transition was believed to be continuous \cite{LAUT,BHA}, but, as
simulations in larger lattices became accessible to computer
resources, the onset of metastabilities revealed the first
order nature of this transition \cite{VIC,ROB,UBER}.

 It was suggested some time ago \cite{BHANOT} that
the order of the PT could be altered when using an extended Wilson
action, including a term proportional to the plaquette squared.
\be
S_{\rm {ext}} = - \sum_{P} [ \beta \cos \theta_{\rm P} + \gamma \cos 2\theta_{\rm P} ]
\label{EXT}
\ee

It was then found \cite{EVER} that for positive values of
$\gamma$ the first order signatures appear for lattices
smaller than the ones needed to state the first order nature
of the PT with the Wilson action ($\gamma=0$). In this way,
$\gamma$ could be used as a parameter to reinforce (positive
$\gamma$) or weaken (negative $\gamma$) the transition, and some
authors \cite{EVER,SPH} conjecture about the existence of a tricritical point
at some negative $\gamma$ value where the order of the transition
changes, becoming a continuous one.

However, numerical simulations showed that metastabilities appear
for negative values of $\gamma$ as well. At this stage, it was
pointed out that one usually works on the lattice with a toroidal
topology, and it was questioned whether or not these metastabilities
survive in the thermodynamic limit, or they are rather  lattice
artifacts due to the toroidal topology \cite{LN}.
The closed monopole loops appearing on the torus were initially
supposed to be responsible for such double peak structures. This
hypothesis led authors to work on lattices homotopic to the sphere, since in
those lattices all monopole loops can be contracted to a point  \cite{SPH}
\cite{LN} \cite{Baig}. In fact, the authors of \cite{SPH}, working on the 
hyper-surface of a 5D cube, which is homotopic to the sphere, do not observe 
any signal of metastability.

This situation is quite disturbing for the lattice community. One would
expect that the topology of the lattice does not affect the physics
of the model, since its contribution behaves as a surface term,
which vanishes in the thermodynamical limit.

To shed some light on this problem, we have studied numerically
the extended Wilson action for several values of the
parameter $\gamma$, both on the torus and, following \cite{SPH}, on the sphere.
On the torus we run simulations up to lattice sizes $L$=24, finding
a clear first order phase transition. On the sphere we work at $\gamma=0$
and at $\gamma=-0.2$, and we, as Jersak et al. \cite{SPH}, do not see any 
two  peak signals 
when we measure in the lattice sizes investigated by them, but when we go to 
larger sizes, we find that the behaviour of the transition is that expected  
for a first order one \cite{LAF}, where for
small lattices $\nu$, being lower than $1/2$, is larger than the
first order value $1/4$, but  approaches  that  value monotonically as the 
lattice size is increased.

We have worked with the plaquette energy defined as
\be
E = \frac{1}{N_{\rm P}}\langle \sum_{\rm P}\cos \theta_{\rm P}\rangle
\label{ENERGY}
\ee
and the specific heat
\be
C_{\rm v} = \frac{\partial}{\partial\beta}E
\label{CSPE}
\ee
where $N_{\rm P}$ stands for the number of plaquettes of the system.

Similar quantities, defined with respect to the term $\cos 2\theta_{\rm P}$, 
have been measured, but they being highly correlated with the previous ones and
their behaviour being qualitatively identical, their results have not been 
reported.

On the Torus, $N_{\rm P}= 6L^4$. On the sphere, the number of plaquettes has a
less simple expression, and can be computed as a function of $N$, the number
of points in one of its $5$ dimensions, as $N_{\rm P}= 60(N-1)^4+20(N-1)^2$. In
this case, the system is
not homogeneous and $N_{\rm P}$ is not proportional to the number of points
on the four dimensional surface, which is $N^5-(N-2)^5$, some points having a 
number of surrounding plaquettes less than the possible maximum $12$, 
as opposed to what happens on the torus . In order to allow comparison, we 
define $L_{\rm {eff}}=(\frac{N_{\rm P}}{6})^{1/4}$.

If the transition is first order, $C_{\rm v}$ must scale at the transition
point as $N_{\rm P}$ in both topologies.

We simulate the subgroup $Z(1024) \subset U(1)$, and
we also ran simulations at some points
with the whole $U(1)$ group, the results with both groups being fully 
compatible for both the spherical and the toroidal topologies.

In order to check the goodness of the simulation, we have used the 
Schwinger-Dyson equations on the lattice \cite{SchDy}, 
which allows us to extract $\beta,\gamma$ from the Monte Carlo data,
the value of the input couplings having been recovered from the simulations on 
both the toroidal and 
spherical topologies.

We intend to give in this letter a schematic presentation of our results.
A complete account of our data will be given in a future
paper \cite{FUTURO}.

\section{Results for the toroidal topology}

We have studied the model on the torus in order to check whether
or not the double peak structures observed in the small volumes survive
in the thermodynamical limit. The smallest lattice we use is $L$=6 and 
the largest one is $L$=24. We update by means of a standard Metropolis
algorithm. The statistics range from 
$\approx 8 \times 10^5$ MC iterations for the smallest lattices ($L$=6,8,12) 
to $ \approx 1.4 \times 10^6$ for the largest ones 
($L$=16,20,24). The autocorrelation time for the energy ranges from
$O(10^2)$ to $O(10^3)$.

For every value of $\gamma = -0.1, -0.2, -0.3, -0.4$ we consider, 
we have used the Spectral Density Method \cite{SDM} 
to locate the critical coupling
$\betac(L)$ at the maximum of the specific heat peak. 

We carried out the simulations on the RTNN machine consisting of 32 Pentium
Pro processors, the total CPU time employed being the equivalent of $4$ 
Pentium Pro 
years.

\begin{figure}[h]
\begin{center}
\epsfig{figure=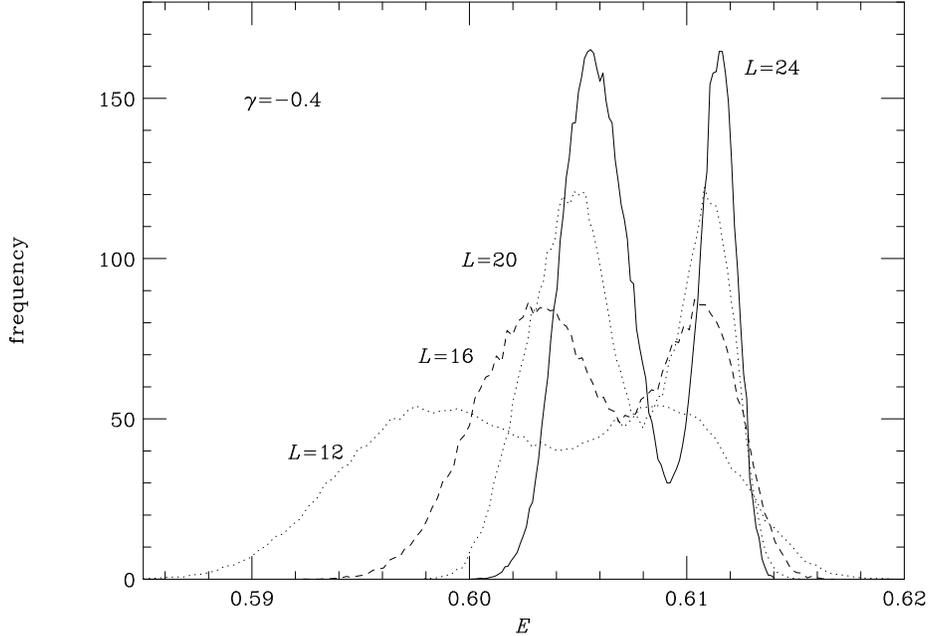,angle=90,width=350pt}
\caption{\small {Plaquette energy distribution measured
at C$^{\mathrm{max}}_v(L)$ for $L$=12,16,20,24
at $\gamma = -0.4$}}
\label{HISTOG}
\end{center}
\end{figure}     

\begin{figure}[h]
\begin{center}
\epsfig{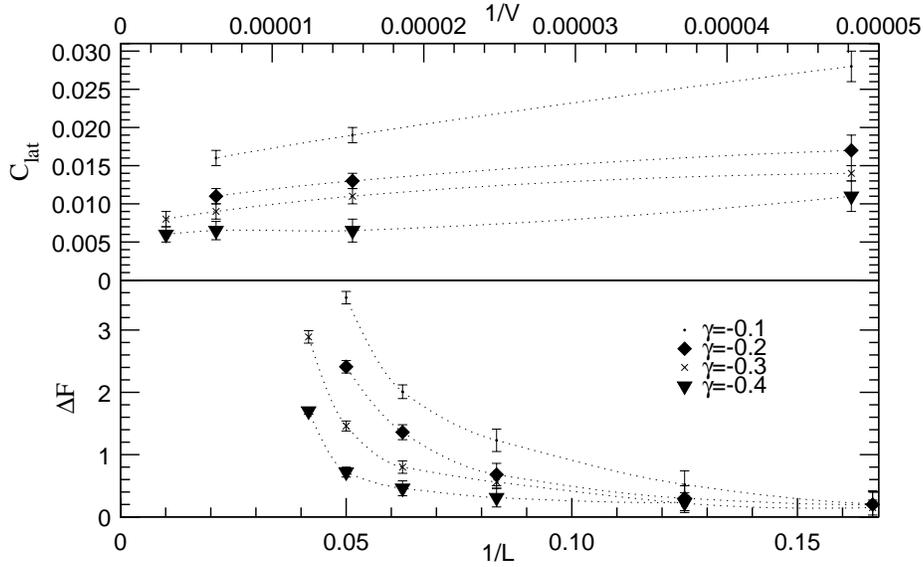}
\caption{\small {Latent heat (upper plot)
and $\Delta F$ (lower plot) for the different $\gamma$ values.}}
\label{LATGAP}
\end{center}
\end{figure}     

We find that the two-state signal persists for all volumes
we consider at all $\gamma$ values.
We plot in figure \ref{HISTOG} the histograms for the
plaquette energy at $\gamma = -0.4$.
The transition shows an increasing weakness as we go to more
negative $\gamma$ values. The double peak structure is clearly
observed in $L$=6 at $\gamma=-0.1$, while at $\gamma = -0.4$ one has to go to 
$L$=12 to observe an equivalent signal.

From the energy distributions,
we measure the latent heat through a cubic spline fit of the peaks. The
result is shown in figure \ref{LATGAP} (upper figure).
The latent heat can be safely extrapolated
to a value different from zero in the thermodynamical limit.

\begin{table}[h]
{

\begin{center}
{
\scriptsize{
\begin{tabular}{|c||c|c|c||c|c|c|} \hline
\multicolumn{4}{|c||}{$\gamma = -0.1$} &  \multicolumn{3}{|c|}{$\gamma = -0.2$} \\ \hline
$L$ & $\beta_{\rm {sim}}$ & $\betac(L)$ & $\nu_{\rm {eff}}$ 
&$\beta_{\rm {sim}}$ &$\betac(L)$  &$\nu_{\rm {eff}}$  \\ \hline
$6$ &1.0720 &1.0716(2)   &-        
 &1.1460 &1.1452(2)   &-                   \\ \hline
$8$ &1.0784  &1.0786(2)  &0.324(13)      
 &1.1535  &1.1539(2) &0.342(16)               \\ \hline
$12$ &1.0820 &1.0818(1)  &0.323(12)     
 &1.1582 &1.1582(2)  &0.352(21)                  \\ \hline
$16$ &1.08278 &1.0827(1)  &0.288(17)      
 &1.15935 &1.1593(1) &0.302(14)                  \\ \hline
$20$ &1.0833 &1.0833(1)  &0.270(16)      
 &1.1599 &1.1599(1) &0.292(18)                  \\ \hline
\hline
\multicolumn{4}{|c||}{$\gamma = -0.3$} &  \multicolumn{3}{|c|}{$\gamma = -0.4$} \\ \hline
$L$&$\beta_{\rm {sim}}$ &$\betac(L)$  &$\nu_{\rm {eff}}$ 
&$\beta_{\rm {sim}}$ &$\betac(L)$  &$\nu_{\rm {eff}}$  \\ \hline
$6$&1.2255   &1.2237(4)  &-      
   &1.3090  &1.3082(4)  &-                   \\ \hline
$8$&1.2344   &1.2340(1) &0.359(11)      
   &1.3192 &1.3194(3)  &0.374(18)               \\ \hline
$12$&1.2395   &1.2395(2)  &0.344(12)     
   &1.3258 &1.3259(1)  &0.372(15)                  \\ \hline
$16$ &1.2410  &1.2410(1)  &0.335(14)      
 &1.32775 &1.3278(1)  &0.360(12)                  \\ \hline
$20$ &1.2416  &1.24156(5)  &0.321(17)      
  &1.3285 &1.3284(1)  &0.313(21)                  \\ \hline
$24$ &1.2417  &1.24162(5)   &0.282(13)      
   &1.3286 &1.3287(1) &0.270(15)                  \\ \hline

\end{tabular}
}
}
\end{center}

}
\caption[a]{\footnotesize{Results obtained for the toroidal topology.}}
\protect\label{TABLA_TORO}

\end{table}

In table $1$ we quote, for the different $\gamma$ values on the torus,
the value of $\beta$ at which we have simulated, and the  $\betac(L)$ obtained
from the maximum of the specific heat using the Spectral Density Method.

For every lattice size,
we have  measured the position of the nearby zero of the partition
function closest to the real axis \cite{ZERO}. The imaginary part of that
value is known to scale as $L^{-1/\nu}$. Following that, we calculated 
an effective $\nu$ exponent between consecutive lattice sizes.
We see that $\nu_{\rm {eff}}$ goes asymptotically
to $1/4$, but for large negative $\gamma$ this value is 
attained for increasingly large L, so evidencing a weaker transition, but
still first order for all $\gamma$ values considered.

From the energy distributions, we can measure an useful 
quantity in order to determine the order of the phase transition, i.e., the 
free energy gap $\Delta F$, which is the difference between the minima
and the local maximum of the free energy \cite{FREE}.
We use the spectral density method to get, from the measured histograms,
a new histogram where both peaks have equal height. We take
the logarithm of those histograms and measure the energy gap.
If the transition is first order, that gap has to be more pronounced as
we go to larger $L$, while it has to stay constant if the transition is
second order.

In figure \ref{LATGAP} (lower plot) we show the behaviour of the energy
gap for the different $\gamma$ values. The gap $\Delta F$ grows
up drastically for all $\gamma$ values, supporting the first order nature
of the phase transition. 
The value of $L$ at which $\Delta F$ starts growing is certainly larger
as the value of $\gamma$ is more negative, revealing the increasing weakness
of the transition as $\gamma$ gets more negative, but there is no suggestion
of the existence of a tricritical point at finite $\gamma$.
This is in agreement with what
one would expect from the behaviour of  $\nu_{\rm {eff}}$. 
Also in this case, a 
pseudo plateau is present for $\Delta F$, larger for larger negative
$\gamma$, and if small lattices are used, that could be interpreted as
a second order behaviour. It should be also noticed that $\Delta F$
scales for the largest lattices as L$^{\rm {d-1}}$, as expected in a 
first order phase transition \cite{FREE}.

In figure \ref{NCV} we plot $C_{\rm v}$ for the torus. In the $x$ axis we plot
the plaquette number $N_{\rm P}$. With this scale, a straight line dependence 
means that
$C_{\rm v}$ scales as the volume, and then $\nu=1/4$. We superimpose a linear 
fit
to the three last points, which is very good, but would not work at smaller
sizes, a behaviour which uses to appear in the
transient region of weak first order phase transitions, somehow preceding
 the onset of the true transition \cite{LAF,ISAF,SU2H,ONAF}. 

\section{Results for the spherical topology}

\begin{figure}[t]
\begin{center}
\epsfig{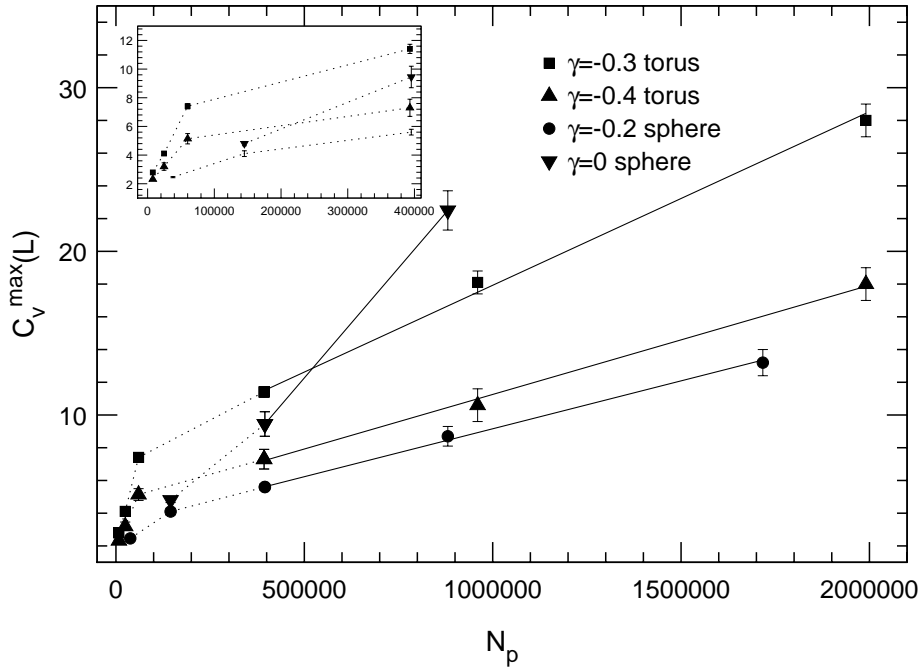}
\caption{$C^{\mathrm{max}}_v(L)$ as a function of the plaquette
number for the sphere at $\gamma=0,-0.2$ and for the torus at
$\gamma=-0.3,-0.4$. For the sphere at $\gamma=0$ we plot 
$C^{\mathrm{max}}_v(L)/2$ for the clarity of the graphic's sake.}
\label{NCV}
\end{center}
\end{figure}     

The first order nature of the deconfinement transition
having been stated for the torus,
we follow Jersak et al.\cite{SPH} and work on the 4D surface of a 5D cube to 
check whether the two-state signal does disappear for that topology.

Based on the torus experience, where we have learned that the behaviour of the
system is similar for combinations of decreasing  $\gamma$ and increasing 
volume, we have chosen to work 
at $\gamma = 0$ and at $\gamma = -0.2$ in lattices ranging from N=6 to N=14.
In order to be sure that no topologically induced large metastabilities are 
present, we have run for larger lattices two independent 
simulations, starting from cold and hot configurations, the results  being 
unaffected by the initial configuration.
We discarded around 20$\%$ of the statistics for thermalization.

We have also measured the position of the first Fisher zero, and computed
a  $\nu_{\rm {eff}}$ in the same way as we did for the toroidal topology.
The results, together with the autocorrelation time ($\tau$) for the energy, 
and the statistics in number of $\tau$, $N_\tau$, are reported in table $2$.

\begin{figure}[t]
\begin{center}
\epsfig{figure=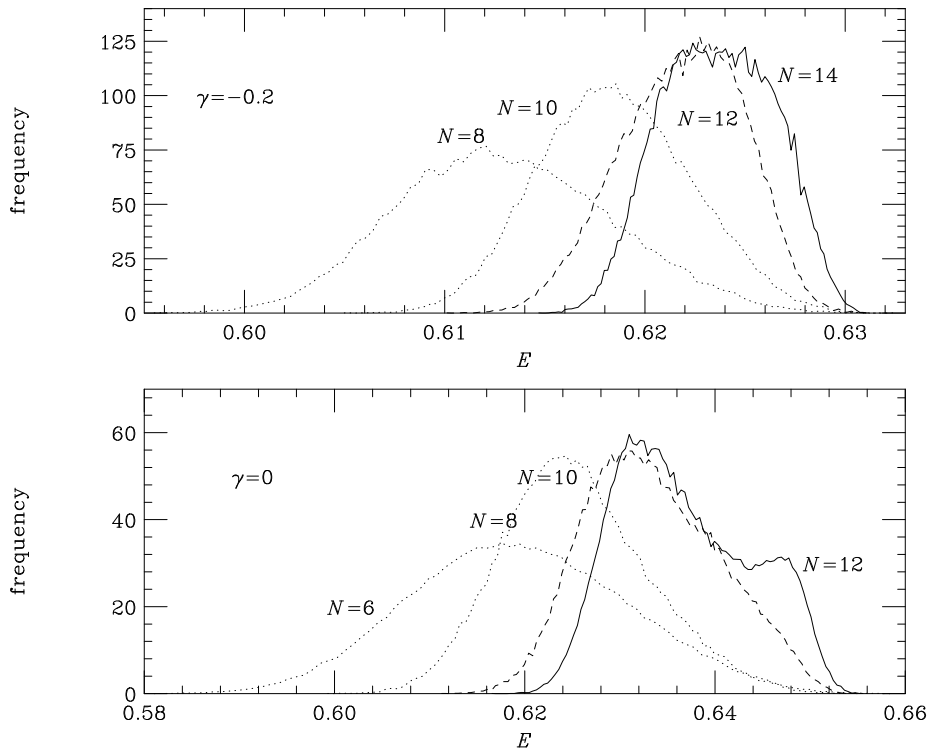,angle=90,width=350pt}
\caption{Plaquette energy distributions in the spherical
lattice at $\gamma=0,-0.2$ at the simulated coupling.}
\label{HESFERA}
\end{center}
\end{figure}

At $\gamma = 0$, (Figure \ref{HESFERA}, lower part), we do not find evidence 
for double peak structures up to $N$=8. However, for $N$=10 the
Energy distribution presents deviations from a simple gaussian behaviour.
The onset of a double-peaked distribution occurs in
$N$=12. 
The values for  $\nu_{\rm {eff}}$ in table \ref{TABLA_ESFERA} show a trend 
towards $1/d$, as expected in a first order phase
transition. The behaviour of the specific heat, proportional to
$N_{\rm P}$ for larger lattices (see figure \ref{NCV}) supports the first order
too.

At $\gamma = -0.2$, (Figure \ref{HESFERA}, upper part), the transition turns 
out to be much weaker than could
be expected from the results obtained with the toroidal topology. In general, a
stable sharp double peak structure can only be observed when the lattice
size is much larger than the correlation length at the critical point.
We do not observe such  signals up to $N$=14, yet at  $N$=14
the distribution is distinctly non-gaussian, and moreover, its width is 
practically identical to the one at $N$=12, which means that $C_{\rm v}$ 
scales 
between both lattices as the volume, or equivalently, that $\nu\approx 1/4$.  

Also the results for  $\nu_{\rm {eff}}$ in table \ref{TABLA_ESFERA} show a 
clear
trend towards the first order value, similar to that shown on the torus in 
table \ref{TABLA_TORO}.
The behaviour of the specific heat,  
(figure \ref{NCV}) is almost compatible with $C_v\approx N_{\rm P}$, 
as expected for a
first order phase transition near the asymptotic region.

In view of all this, our hypothesis is that the two peaks of the energy 
distribution
are too close to be discerned up to $N=14$, but the results for the scaling
of the specific heat start to be significant from $N=12$ on.
From the behaviour observed at $\gamma = 0$ we hope that the splitting of
the peaks in the energy distribution shall be visible for $N=16$.
We are running simulations in that lattice to support this
conjecture \cite{FUTURO}.

\begin{table}[h]
{

\begin{center}
{
\scriptsize{
\begin{tabular}{|c|c||c|c|c|c|c|} \hline
\multicolumn{7}{|c|}{$\gamma = 0$} \\  \hline
$N$  &$L_{\rm {eff}}$ &$\beta_{\rm {sim}}$ &$\tau$ &$N_\tau$  & $\betac(L)$ & $\nu_{\rm {eff}}$ \\ \hline
$6$  &$8.921$    &1.0128 &240   &8300 &1.01340(1)     &-          \\ \hline
$8$  &$12.469$   &1.0120 &400   &2300 &1.0128(2)      &0.296(42)    \\ \hline
$10$ &$16.021$   &1.0120 &780   &1200 &1.01212(3)     &0.264(23)    \\ \hline
$12$ &$19.574$   &1.0119 &850   &1150  &1.01194(2)     &0.240(9)    \\ \hline
\hline
\multicolumn{7}{|c|}{$\gamma = -0.2$} \\  \hline
$N$  &$L_{\rm {eff}}$ &$\beta_{\rm {sim}}$ &$\tau$&$N_\tau$  & $\betac(L)$ & $\nu_{\rm {eff}}$ \\ \hline
$6$  &$8.921 $   &1.1587 &160   &2000 &1.1587(4)     &-          \\ \hline
$8$  &$12.469$   &1.1597 &510   &1000 &1.1603(2)     &-          \\ \hline
$10$ &$16.021$   &1.1602 &680   &1500 &1.1604(2)     &0.376(25)    \\ \hline
$12$ &$19.574$   &1.1604 &820   &1200 &1.1602(1)     &0.289(23)    \\ \hline
$14$ &$23.123$   &1.1605 &900   &1100 &1.16048(5)    &0.264(19)    \\ \hline
\end{tabular}
}
}
\end{center}

}
\caption[a]{\footnotesize{Results obtained for the spherical topology.}}
\protect\label{TABLA_ESFERA}

\end{table}

Uncontrolled finite size effects seem to be much larger for the spherical 
topology than for the toroidal one. As an example, in $\gamma = 0$, one has to
go to the surface of a 5D, $N$=12 cube  (which has roughly the same
number of points as a $L$=20 toroidal lattice) to observe signals 
comparable to those in a $L$=8 lattice with toroidal topology.
This could be due to the fact that in the cube surface translational invariance
is lost, and then the
thermodynamical limit is reached only for larger lattices than in the
torus topology. This loss of translational invariance, which is more important
for smaller lattices, can be somewhat
alleviated introducing appropriate weight factors in the edges \cite{SPH},
which we did not consider, since such details are not expected to affect the 
order of the phase transition.

\section{Conclusions and outlook}

We have studied the 4D pure compact U(1) gauge theory 
with the extended Wilson action
on the torus and on the sphere. On the torus the evidence supports
the idea of a first order phase transition between the Confined phase and
the Coulomb one, as stated some years ago using RG techniques \cite{HASEN}. 
On the sphere,
our preliminary results point to a first order transition at $\gamma=0$.
At $\gamma=-0.2$ the transition seems to be weaker, but still
first order. This statement is supported by the behaviour of $\nu_{\rm {eff}}$,
which goes to $1/4$ in both topologies, with numerical coincidence at 
corresponding sizes (see tables 1 and 2), and by an
incipient double peak structure at N=14. The behaviour of the
specific heat, and the scaling of the Fisher zeros,
are hardly compatible with a second order phase transition.
This result has to be confirmed in larger lattices. The evidence produced in
favour of the first order character supports the folklore
that both, the torus and the sphere, should exhibit
the same behaviour in the thermodynamical limit.

Also, one should be aware of the fact that
in weak first order phase transitions
an exponent  $\nu_{\rm {eff}}$ between $1/d$ and $1/2$ appears 
during its transitory
region \cite{BHANOT,LAF,ISAF,SU2H,ONAF} and hence it should not be
surprising that this be the behaviour exhibited by this model.

\section*{Acknowledgments}
We have benefited from comments and discussions with L.A. Fern\'andez. 
We have carried out our simulations on dedicated Pentium Pro machines 
(RTNN project).
We thank CICyT (contract AEN97-1708) for partial
financial support. I. Campos is a Spanish MEC fellow.


\newpage



\begin{thebibliography}{99}

\bibitem{LAUT}
B. Lautrup and M. Nauenberg
{\sl Phys. Lett.} {\bf B95, p 63} (1980)
\bibitem{BHA}
G. Bhanot
{\sl Phys. Rev.} {\bf D24, p 461} (1981)
\bibitem{VIC}
V. Azcoiti, G. di Carlo and A. Grillo
{\sl Phys. Lett.} {\bf B238, p 355} (1990)
\bibitem{ROB}
L.A. Fern\'andez, A. Mu\~noz-Sudupe, R. Petronzio and A. Taranc\'on. 
{\sl Phys. Lett.} {\bf B267, p 100} (1991)
\bibitem{UBER}
G. Bhanot, T. Lippert, K. Schilling and P. Ueberholz
{\sl Nuc. Phys.} {\bf B378, p 633} (1992)
\bibitem{BHANOT}
G. Bhanot
{\sl Nuc. Phys.} {\bf B205, p 168} (1982)
\bibitem{EVER}
H.G. Evertz, J. Jersak, T. Neuhaus and P.M. Zerwas
{\sl Nuc. Phys.} {\bf B251 p 279} (1985)
\bibitem{SPH}
J. Jersak, C.B. Lang and T. Neuhaus
{\sl Phys. Rev. Lett.} {\bf 77, p 1933} (1996)
\bibitem{LN}
C.B. Lang and T. Neuhaus
{\sl Nuc. Phys.} {\bf B431 p 119} (1994)
\bibitem{Baig}
M. Baig and H. Fort
{\sl Phys. Lett.} {\bf B332, p 428} (1994)
\bibitem{SchDy}
M. Falcioni {\sl et al.},
{\sl Nucl. Phys.} {\bf B265, p 187} (1986)
\bibitem{FUTURO}
I. Campos, A. Cruz and A. Taranc\'on, work in preparation.
\bibitem{SDM}
A. M. Ferrenberg and R. Swendsen.
{\sl Phys. Rev. Lett.} {\bf 61, p 2635}(1988)
\bibitem{ZERO}
C. N. Yang and T. D. Lee.
{\sl Phys. Rev.} {\bf 87, p 404} (1952)
\bibitem{FREE}
J. Lee and J.M. Kosterlitz
{\sl Phys. Rev. Lett.} {\bf 65, p 137} (1990)
\bibitem{LAF}
L.A. Fern\'andez, M.P. Lombardo, J.J. Ruiz-Lorenzo and A. Taranc\'on.
{\sl Phys. Lett.} {\bf B277, p 485} (1992)
\bibitem{ISAF}
J.L. Alonso, J.M. Carmona, J. Clemente, L.A. Fern\'andez, D. I\~niguez,
A. Taranc\'on and C.L. Ullod.
{\sl Phys. Lett.} {\bf B376, p 148} (1996)                  
\bibitem{SU2H}
I. Campos
{\sl Nuc. Phys.} {\bf B514, p 336} (1998) 
\bibitem{ONAF}
H.G. Ballesteros, J.M. Carmona, L.A. Fern\'andez and A. Taranc\'on
{\sl hep-lat:/9707030, to appear in Phys. Lett. B}
\bibitem{HASEN}
A. Hasenfratz
{\sl Phys. Lett.} {\bf B201, p 492} (1988)
\end{thebibliography}
\end{document}